# Direct Observation and Quantitative Measurement of OH Radical Desorption During the Ultraviolet Photolysis of Liquid Nonanoic Acid

Naoki Numadate,[†,*] Shota Saito,[†] Yuki Nojima,[‡] Taka-aki Ishibashi,[‡] Shinichi Enami,[¶] and Tetsuya Hama[†,**]

[†]*Komaba Institute for Science and Department of Basic Science, The University of Tokyo, 4-6-1 Komaba, Meguro, Tokyo 153-8505, Japan*

[‡]*Department of Chemistry, Graduate School of Science and Technology, University of Tsukuba, Tsukuba 305-8571, Japan*

[¶]*National Institute for Environmental Studies, Tsukuba 305-8506, Japan*

AUTHOR INFORMATION

**Corresponding Authors**

*Naoki Numadate and **Tetsuya Hama

E-mail: numadate-naoki@g.ecc.u-tokyo.ac.jp and hamatetsuya@g.ecc.u-tokyo.ac.jp

Phone: +81-3-5452-6288

ABSTRACT

Ultraviolet (UV) photolysis of fatty acid surfactants such as nonanoic acid (NA)—which cover the surfaces of atmospheric liquid aerosols and are found in the oceans—has recently been suggested as a source of hydroxyl (OH) radicals in the troposphere. We used laser-induced fluorescence to directly observe OH radicals desorbed from the surface of neat liquid NA as a primary photoproduct following 213 nm irradiation. The upper limit of photoreaction cross section for the OH radical desorption was estimated to be $9.0(4.1) \times 10^{-22}$ cm$^2$, which is only $1.2 \pm 0.8$ percent of the photoreaction cross section established for the photolysis of gas-phase acetic acid monomers. Vibrational sum-frequency generation spectroscopy for liquid NA revealed the hydrogen-bonded, cyclic, dimer structure of the NA molecules at the liquid surface. This dimerization can inhibit the formation of OH radicals and lead the present low photochemical reactivity of liquid NA.

**TOC GRAPHICS**

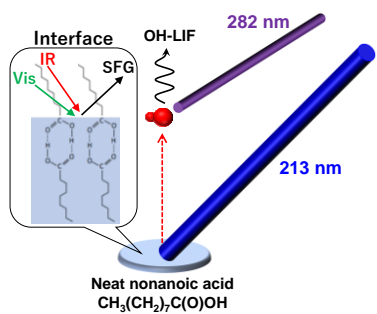

Quantitative studies of photo- and radical chemistry at the interfaces of atmospheric aerosols and at the sea surface are particularly important for further understanding local and global climate change. This is because these interfaces are substantial natural reaction fields[1–3] and significantly affect Earth's energy budget.[4] Photochemical reactions of organic surfactants such as fatty acids—which cover the surfaces of atmospheric aerosols and the sea—have attracted much attention. For example, Rossignol *et al*. used switchable reagent ionization time-of-flight mass spectrometry to find that a monolayer of nonanoic acid (NA) at the air–water interface absorbs ultraviolet (UV) light in the 280–330 nm region and then produces complex, gas-phase, organic compounds such as $C_8$ and $C_9$ aldehydes, which eventually act as aerosol precursors.[2] Although the main absorption band of neat NA or its solutions has a maximum at around 210 nm via a singlet–singlet ($S_0$–$S_1$) transition (Fig. S1),[2] Rossignol *et al*. proposed two possible photochemical reaction mechanisms induced by a weak singlet–triplet ($S_0$–$T_1$) transition at 280–330 nm; i.e., an intermolecular Norrish type-II reaction (hydrogen abstraction from NA by adjacent photoexcited NA) and hydroxyl (OH) radical production through photodissociation (C–O bond fission).[2] The latter would significantly affect atmospheric chemistry, because OH radicals are important oxidants in the atmosphere.[5,6]

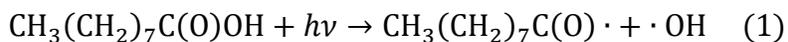

$$\mathrm{CH_3(CH_2)_7C(O)OH} + h\nu \rightarrow \mathrm{CH_3(CH_2)_7C(O)\cdot} + \cdot \mathrm{OH} \quad (1)$$

Xiao *et al*. subsequently studied photochemical reactions of NA molecules using first-principles density functional theory.[7] Their theoretical results suggested that the $S_0$–$S_1$ transition is important to the $S_0$–$T_1$ photoexcitation of isolated NA at around 270 nm, which is followed by $S_1 \rightarrow T_1$ intersystem crossing. They also proposed that the formation of $C_9$ aldehydes is initiated by intramolecular $T_1$ C–O bond fission of NA, generating acyl and OH radicals, rather than by the intermolecular Norrish type-II reaction; the $T_1$ C–C bond fission, generating octyl and carboxyl radicals, is the initiation process for the formation of $C_8$ aldehydes.[7]

Although the above experimental and theoretical studies advocate the importance of direct detection of primary photoproducts, especially OH radicals, to understand the photochemistry of NA, previous experimental approaches using mass spectrometry with continuous UV irradiation are not suitable for the detection of highly reactive species such as OH radicals.[2] Consequently, the mechanisms of the photochemical reactions of fatty acid surfactants such as NA remain poorly understood even in the neat liquid phase, let alone in a monolayer at the air–water interface. An alternative technique for detecting radical products is highly desirable to improve our fundamental understanding of the photochemistry of liquid fatty acids, which would eventually allow elucidation of their impact on the troposphere.

This study developed a new experimental apparatus for the UV laser photolysis of liquids by laser-induced fluorescence (LIF; Figs 1 and S2; see the Experimental Methods and Supporting Information for details). Our new approach enables direct detection of OH radicals photodesorbed from the surface of a liquid organic sample. As a first step, this study focused on the 213 nm photoexcitation of neat liquid NA via the $S_0$–$S_1$ transition, because liquid NA has a large absorption cross section of about $10^{-19}$ cm$^2$ at around 210 nm (Fig. S1 and ref. 8). We demonstrated that OH radicals are desorbed as a primary photoproduct following 213 nm excitation of liquid NA. We also quantified the photoreaction cross section of OH radical desorption by comparing the 213 nm photolyses of liquid NA and gas-phase acetic acid (AA). Heterodyne-detected (HD) vibrational sum-frequency generation (VSFG) spectroscopy provided additional information about the mechanism of the photochemical reactions of liquid NA. Considering both the LIF and HD–VSFG results enabled us to discuss the relationship between the surface structure and the photochemical reactivity of liquid NA.

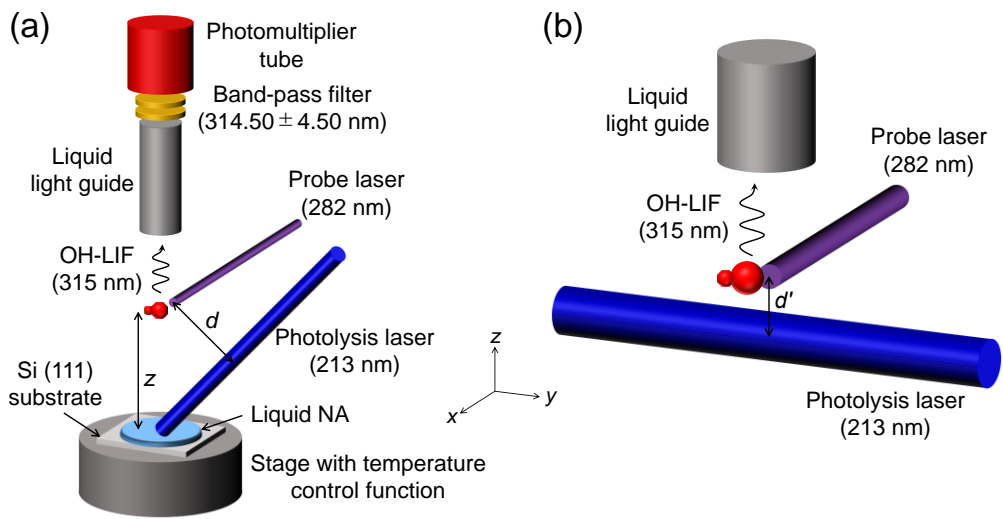

**Figure 1.** Schematic diagrams of 213 nm photolysis of liquid- and gas-phase organic molecules and OH radical probing (see the Experimental Methods for details). (a) Liquid-phase photolysis used a laser incident on the sample at 45° to its surface normal. (b) Gas-phase photolysis used two lasers crossing in a gas-filled chamber. Laser-induced fluorescence via the $A^2\Sigma$–$X^2\Pi$ transition of OH radicals was detected using a photomultiplier tube through a liquid light guide and two band-pass filters. $z$ indicates the distance between the liquid surface and the center of the dye laser. $d$ ($= z/\sqrt{2}$) and $d'$ indicate the shortest distances between the centers of the two lasers.

*Laser delay dependence*. Figure 2 shows normalized OH-LIF intensities with respect to delay time between the photolysis and probe lasers for the liquid- and gas-phase experiments illustrated in Fig. 1a and b (Figs S3 and S4 in the Supporting Information give further details of the data acquisition). The intensities during photolysis of liquid NA (red circles in Fig. 2) were measured using different detection points (3 mm ≤ $z$ ≤ 7 mm), where $z$ represents the distance between the liquid surface and the center of the dye laser (Fig. 1). The peak delay time for the OH-LIF intensity increased as $z$ increased from 3 to 7 mm. This strongly suggests that the OH radicals desorbed from the liquid surface. Nevertheless, gaseous NA molecules were also present in the chamber, considering the vapor pressure of liquid NA (approximately 0.2 Pa) at room temperature.[9] Therefore, we separately conducted 213 nm photolysis of gas-phase NA using two crossed lasers (as shown in Fig. 1b) to evaluate the contribution of gas-phase photolysis of NA to the OH-LIF signals in the liquid experiment. The OH-LIF intensities from the gas-phase photolysis (black vertical bars in Fig. 2) were measured at different detection points (2 mm ≤ $d'$ ≤ 5 mm in Fig. 1b). The distance between the photolysis and probe lasers was approximately identical to that in the liquid-phase experiment (i.e., $d \approx d'$ in Fig. 1a and b). The beam sizes and pulse energies of the two lasers in the gas-phase experiment were also identical to those in the liquid-phase experiment. When the two lasers were close to each other ($d' = 2$ mm), an OH-LIF signal due to gas-phase NA photolysis was observed (Fig. 2). However, the OH radicals became barely detectable upon increasing $d'$ to 5 mm. The difference between the OH-LIF intensities measured in the liquid and gas phases should correspond to the amount of OH radicals desorbed from the liquid NA, which was much larger than that from the gas-phase NA photolysis. These results confirm that the dominant part of the OH-LIF signals during the liquid-phase experiment originated from OH radicals desorbed from the liquid NA surface.

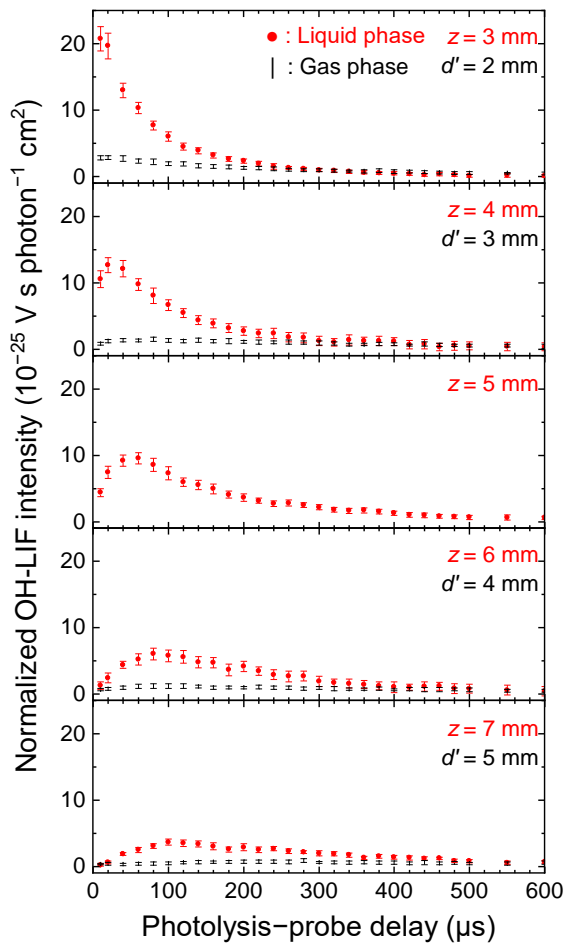

**Figure 2.** Normalized OH-LIF intensities with respect to the delay time between the photolysis and probe lasers. Red circles and black vertical bars indicate the intensities measured in the liquid- and gas-phase experiments, respectively, at 80 Pa with nitrogen as a buffer gas. Each data point represents the average of 160 laser shots. The OH-LIF intensities are normalized by the photon fluences of the photolysis laser: $3.1 \times 10^{15}$ photon cm$^{-2}$ for the liquid-phase experiments and $4.4 \times 10^{15}$ photon cm$^{-2}$ for the gas-phase experiments. Both experiments were performed under similar conditions regarding the distance between the centers of the two lasers ($d = z/\sqrt{2} \cong d'$). The error bars indicate 1σ statistical errors.

*Single-shot measurement.* We also conducted single-shot experiments to reveal the origin of the photodesorbed OH radicals; i.e., whether they were formed directly from liquid NA molecules or from secondary photoproducts accumulated on the liquid surface. Figure 3 shows the time evolution of the OH-LIF signal intensities with respect to the 213 nm laser shots toward the liquid NA at the on- and off-resonance wavelengths for the excitation of OH radicals (Fig. S4). The pulsed probe laser was set to be injected into the chamber during the single-shot experiments. At the on-resonance wavelength (black lines in Fig. 3), the intensities steeply increased immediately after 213 nm irradiation, then remained almost constant during 400 subsequent shots of irradiation in all measurements at repetition rates of 1, 2, 5, and 10 Hz for the two lasers. The uncertainties of the OH-LIF signal intensities were derived from the energy fluctuations of the two lasers. At the off-resonance wavelength, no OH-LIF signals appeared after 213 nm irradiation (gray lines in Fig. 3), which corresponds to the background signal intensities derived from stray light and the scattering of the probe laser. These results mean that the photolysis of NA molecules forms OH radicals as a primary photoproduct, and the desorption of OH radicals by 213 nm photolysis of secondary photoproducts accumulated on the liquid sample can be neglected at the time scale of 0.1–1 s.

The results in Figs 2 and 3 indicate that OH radicals desorbed following the 213 nm photolysis of neat liquid NA.

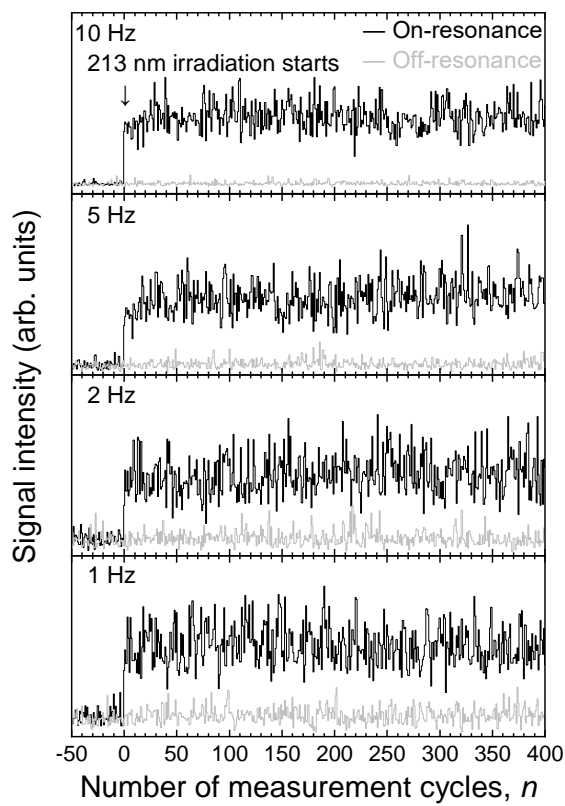

**Figure 3.** Time evolution of the integrated signal intensity with respect to the number of measurement cycles, obtained by single-shot data acquisition at various laser repetition rates during the photolysis of liquid NA. Black and gray lines indicate results measured at the on- and off-resonance wavelengths for OH radical probing, respectively. Data were obtained at 80 Pa using a buffer of nitrogen gas. The 213 nm photolysis laser irradiation of the liquid NA surface commenced at $n = 0$.

*Photoreaction cross section for OH radical desorption.* This study could not accurately determine the quantum yield of OH radical formation owing to a lack of data regarding the absorption cross section of NA molecules at the liquid interface. Hence, we discuss the photochemical reactivity of liquid NA by comparing the OH-LIF signal intensities from the 213 nm photolyses of gas-phase AA and liquid NA, because the absorption cross section and quantum yield of the OH formation have been well studied for gas-phase AA (see Table S1 for details)[10,11]. Here we define the photoreaction cross section as $[OH]/F_{photon}^{213} [X]$ (cm$^2$):

$$[OH] = \sigma_{abs}^{213}(X) \times F_{photon}^{213} \times \phi_{OH}^{213}(X) \times [X], \quad (2)$$

$$\frac{[OH]}{F_{photon}^{213}[X]} = \sigma_{abs}^{213}(X) \times \phi_{OH}^{213}(X). \quad (3)$$

[OH] is the number density of OH radicals in cm$^{-3}$. [X] is the number density of liquid NA molecules [NA] or gas-phase AA molecules [AA] in cm$^{-3}$. $\sigma_{abs}^{213}(X)$ is the absorption cross section at 213 nm for molecule X in cm$^2$. $F_{photon}^{213}$ is the photon fluence of the 213 nm photolysis laser in cm$^{-2}$, and $\phi_{OH}^{213}(X)$ is the quantum yield of OH radicals in the photolysis of molecule X at 213 nm.

First, we obtain a photoreaction cross sections of $7.6(1.8) \times 10^{-20}$ cm$^2$ for gas-phase AA monomers and $7.2(5.3) \times 10^{-21}$ cm$^2$ for gas-phase AA dimers by substituting literature values into Eq. 3 (Table S1 gives details).[10,11] In addition, the photolysis of gas-phase AA was conducted in the crossed-beam configuration (Fig. 1b) at 295 K at a partial pressure of 1.2 Pa for gas-phase AA in a total pressure of 80 Pa with nitrogen as a buffer gas. The number density of gas-phase AA ([AA]) was $2.9 \times 10^{14}$ cm$^{-3}$, and the photon fluence of the photolysis laser was $1.5 \times 10^{14}$ cm$^{-2}$. The relationship between the number density of the formed OH radicals and the OH-LIF signal intensity can be written as follows:

$$[OH] = I_{LIF} \times \varepsilon \times \frac{1}{\eta} \times \frac{2\pi}{\Omega}, \quad (4)$$

where $I_{LIF}$, $\varepsilon$, $\eta$, and $\Omega$ indicate the measured OH-LIF intensity (V s), a conversion factor between the OH-LIF intensity and the number density of OH radicals (V$^{-1}$ s$^{-1}$ cm$^{-3}$), the efficiency of detecting OH radicals, and the solid angle of OH-LIF detection, respectively. Therefore, the photoreaction cross section $[OH]/F_{photon}^{213}[X]$ becomes

$$\frac{[OH]}{F_{photon}^{213}[X]} = \sigma_{abs}^{213}(X) \times \phi_{OH}^{213}(X) = \frac{I_{LIF}}{F_{photon}^{213}[X]} \times \varepsilon \times \frac{1}{\eta} \times \frac{2\pi}{\Omega}. \quad (5)$$

The present experiments had the electronics and optics arranged to make $\varepsilon$ and $\Omega$ common in the liquid- and gas-phase experiments. The detection efficiency, $\eta$, depended on the diffusion of photodesorbed OH radicals and the decrease of their concentration by reactions with each other or with gas-phase AA or NA. In this study, the nitrogen buffer gas was introduced into the chamber up to 80 Pa, and hence the OH radicals during the gas-phase and liquid photolyses were cooled to room temperature due to repeated collisions with the buffer gas. The cooling of the OH radicals is necessary to derive the photoreaction cross section because this makes the diffusion velocities of the OH radicals identical in all the experiments, which enables us to quantitatively compare the photolyses of the gas-phase AA and the liquid NA. In addition, given the low number density of the formed OH radicals (~10$^9$ cm$^{-3}$), OH–OH reactions can be ignored here. Reactions between OH and gas-phase AA or NA can also be ignored for the following reason. The rate constants for the reactions of OH radicals with organic compounds are typically 10$^{-13}$–10$^{-10}$ cm$^3$ molecule$^{-1}$ s$^{-1}$,[14] and the formed OH radicals are considered to be cooled to room temperature by multiple collisions with the nitrogen buffer gas in this study. Overall, the mean free paths of the OH radicals in the gas-phase AA or NA can be roughly estimated to be at least dozens of millimeters, which is much greater than the typical flight lengths of the OH radicals before LIF detection.

For valid comparison of the OH-LIF signal intensity of the gas-phase AA photolysis with that of the liquid-phase NA photolysis, $\eta$ should be the same in both experiments. Hence, the OH-LIF signal intensities measured at $z = 0$ mm and $d' = 0$ mm allow the diffusion of the OH radical products to be neglected. However, it is impossible to measure the intensity at $z = 0$ mm due to the finite beam size of the probe laser. Therefore, we investigated the dependence of OH-LIF intensities during liquid NA photolysis on the detection position $z$ (Fig. 4). The OH-LIF signal intensity became almost constant when the distance from the liquid surface to the probe laser ($z$) was less than 3 mm. This was not due to saturation of the detector, because the raw OH-LIF intensity measured during the gas-phase AA photolysis, before normalization by the photon fluence of the photolysis laser, was approximately two or three times larger than that measured during the liquid-phase NA photolysis (Table S1). Thus, Fig. 4 indicates that the diffusion of the formed OH radicals was sufficiently suppressed when $z \leq 3$ mm. Assuming that the OH-LIF signal intensity measured at $z = 1.5$ mm was close to that for $z = 0$ mm, the present analysis compares the results of the liquid-phase photolysis of NA at $z = 1.5$ mm with those of the gas-phase photolysis of AA at $d' = 0$ mm. Both were measured at the laser delay time of 10 μs.

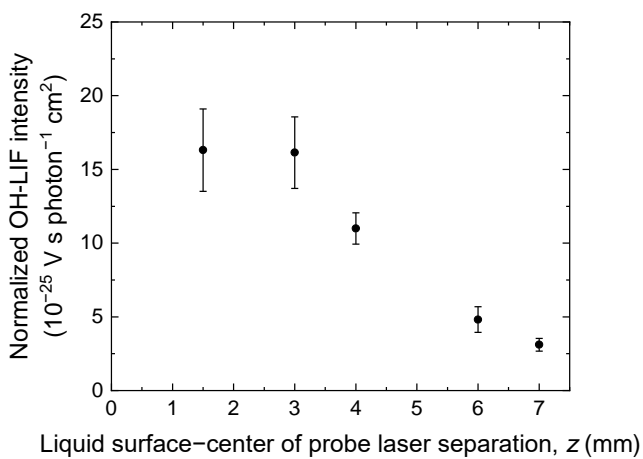

**Figure 4.** Dependence on detection position of OH-LIF intensities during the 213 nm photolysis of liquid NA. The intensities were measured at 80 Pa with nitrogen buffer gas and normalized by the photon fluence of the photolysis laser ($2.5 \times 10^{15}$ photon cm$^{-2}$). Contributions from the background and the photolysis of gas-phase NA are subtracted.

The above discussions lead to the value of $\varepsilon \times \frac{1}{\eta} \times \frac{2\pi}{\Omega}$ being common to both the gas-phase AA and liquid-phase NA photolysis. In the gas-phase AA photolysis, the effect of dimerization on the quantum yield of OH radicals should be considered for the quantitative analysis, because Singleton *et al.* found that the quantum yields of OH radical formation during the 222 nm photolysis of monomer and dimer AA are 0.546 and 0.038, respectively.[10] The molar fractions of monomer and dimer AA under the condition of our experiments are calculated as 0.62 and 0.38, respectively, using the dimerization equilibrium constants given by Chao *et al.*[12] and Crawford *et al.*[13].

To derive the photoreaction cross section $[OH]/F_{photon}^{213}[NA]$, the number density of NA molecules involved in the photochemical reaction [NA] is also required. The bulk density of neat liquid NA is $3.3 \times 10^{21}$ cm$^{-3}$.[15] However, this value is not applicable as [NA], because the optical depth of neat NA at 213 nm is about 10 μm (Fig. S1 and ref. 8). In addition, OH radicals formed at micrometer depths would not desorb, owing to their small mean free path in the bulk.[16] As we cannot determine accurately the region in which the detected OH radicals formed, this study assumes that the OH photoproducts are desorbed only from the first layer of liquid NA. This assumption gives an upper limit for the photoreaction cross section and is useful for the subsequent discussion of the present results. Danov *et al.* expected the molecular area of NA in an aqueous solution to be $2.26 \times 10^{-15}$ cm$^2$,[17] which corresponds to about $4.4 \times 10^{14}$ cm$^{-2}$ on the surface of liquid NA. George *et al.* also reported the saturation of the number density of NA molecules at the water interface to be about $3.2 \times 10^{14}$ cm$^{-2}$.[2,18] Here, if the OH photoproducts are desorbed only from the first layer of liquid NA, the value for [NA] (cm$^{-3}$) corresponds to that for the areal density of NA molecules ($3.2$–$4.4 \times 10^{14}$ cm$^{-3}$), which is the lower limit of [NA]. We thus can use the average of two values ([NA] = $3.8(0.6) \times 10^{14}$ cm$^{-3}$) to obtain the upper limit of the photoreaction cross section for liquid NA as $9.0(4.1) \times 10^{-22}$ cm$^2$ from Eq. 5 using the measured OH-LIF signal intensity and the value of $\varepsilon \times \frac{1}{\eta} \times \frac{2\pi}{\Omega}$ (see Table S1 for details of the parameters).

Consequently, the ratio of the photoreaction cross section of liquid NA to that of gaseous monomer AA is derived to be $1.2(0.8) \times 10^{-2}$. This ratio is the upper limit for the reactivity of liquid NA on the assumption described above. The cross section for the OH radical desorption in the photolysis of liquid NA $[\text{OH}]/F_{\text{photon}}^{213}[\text{NA}]$ would decrease if the OH photoproducts were desorbed from deeper regions, which means an increase in the value of [NA]. Therefore, it is concluded that the liquid-phase photolysis of NA forms fewer OH radicals than the gas-phase photolysis of AA. Table S1 gives the values and errors of each parameter used in this study.

A possible reason for this small OH radical yield is the formation of cyclic dimer structures at the liquid NA surface. Tyrode *et al.* used VSFG spectroscopy to find that AA molecules at the surface of an aqueous AA solution form cyclic AA dimers at high AA concentrations.[19] Molecules at the surface of neat liquid NA can also form cyclic NA dimers through strong hydrogen bonding, which might inhibit OH radical formation during the photolysis of liquid NA.

*Sum-frequency generation (SFG) spectroscopy*. We used HD–VSFG spectroscopy to examine the possibility of cyclic dimer formation at the interface of air and neat NA.[20–22] An SFG signal arises only in media with no centrosymmetry such as interfaces, because SFG is a second-order nonlinear optical process. Therefore, VSFG spectroscopy can provide selective information on molecular orientations at the interface. Unlike infrared (IR) or Raman spectroscopy, HD–VSFG spectroscopy can distinguish the "up" or "down" orientations of a specific substituent group at the interface.[23–25]

Figure 5A shows the SFG spectrum in the C=O stretching region; the plot of Im $\chi^{(2)}$ is the imaginary part of $\chi^{(2)}$ at the air–NA interface. The imaginary part, where a vibrational resonance appears as an absorptive line, is easily compared with bulk IR and Raman spectra. No vibrational band was observed from the neat NA liquid surface, and a vibrational band was clearly observed at 1714 cm$^{-1}$ from the interface of air and a 1.0 mM aqueous NA solution (Figure S5). There are two possible interpretations for the absence of C=O bands in Fig. 5A. One is a random orientation of NA molecules at the interface, as the SFG signal is generated when molecules have a specific orientation at the interface. The other possible interpretation is that a pair of NA molecules form a cyclic dimer structure when their carboxyl groups are hydrogen bonded facing each other. In this case, C=O stretching modes should be SFG inactive, because the hydrogen-bonded carboxylic acid portion is locally centrosymmetric. We observed SFG spectra in the CH stretching region for the interface between air and neat NA to evaluate which of the two interpretations is more plausible.

Figure 5B shows the Im $\chi^{(2)}$ spectrum in the CH stretching region for the interface between air and neat NA. In contrast to the C=O stretching region, vibrational bands appear clearly. The signs of the CH stretching bands indicate that the NA molecules were oriented with their terminal methyl groups toward the air at the interface.[26] This suggests that the absence of C=O bands in the spectrum of the air–neat NA interface was not due to NA molecules being disoriented, but instead due to the formation of cyclic dimers at the interface.

By comparing the signal intensities of the C=O stretching band between the SFG spectra of the air–1.0 mM aqueous NA and air–neat NA interfaces (Fig. S5 and 5A), an occupancy of NA molecules without forming cyclic dimers (non-dimers) at the air–neat NA interface can be roughly estimated. In this study, the detection limit of the C=O stretching band signal is approximately one-fifth of that obtained from the air–1.0 mM NA aqueous interface (Fig. S5). Assuming that all the NA molecules at the interface between air and 1.0 mM aqueous NA are monomers (non-dimers), the non-detection of the C=O stretching band in Fig. 5(A) suggests that the occupancy of non-dimers is at most 20% at the air–neat NA interface and the majority of NA molecules forms a cyclic dimer-structure, implying a high surface homogeneity of neat liquid NA.

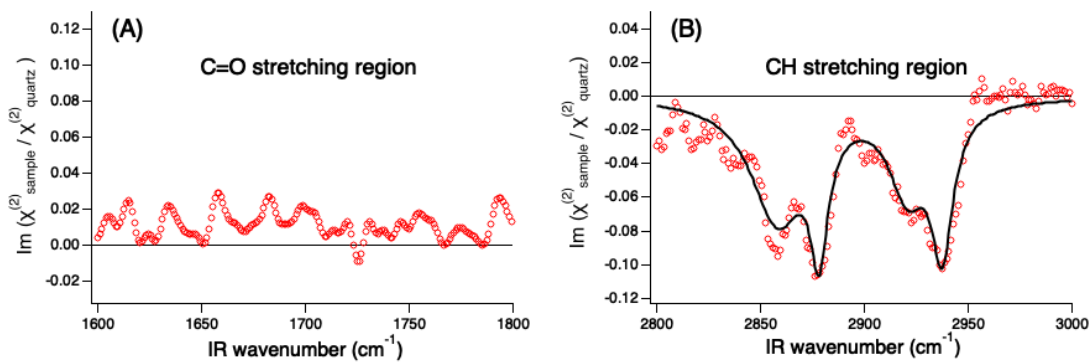

Figure 5. SFG spectra of the air–neat NA interface in the (A) C=O stretching region and the (B) CH stretching region. The black solid line in (B) represents curve fitting by the sum of complex Lorentzian functions. Table S2 lists the obtained fitting parameters.

In this study, we developed new apparatus with the aim of quantitatively studying OH radical desorption during liquid organic photolysis. We succeeded in the direct detection of OH radicals formed by the 213 nm photolysis of neat liquid NA by using a highly sensitive LIF technique. Comparing the results obtained during the photolysis of liquid-phase NA and gas-phase AA, we estimated the photoreaction cross section of OH radical desorption in the 213 nm photolysis of liquid NA to be $9.0(4.1) \times 10^{-22}$ cm$^2$ as an upper limit, which is only $1.2 \pm 0.8$ percent of that for the gas-phase AA monomer. This small value might have been due to the formation of cyclic NA dimers at the liquid surface, as indicated by HD-VSFG spectroscopy of the interface between air and neat NA. Moreover, some OH products may be consumed by reactions with surrounding NA molecules without escaping from the liquid surface, which can reduce the cross section for the OH radical desorption.

The photolysis of aqueous NA, namely the photolysis of hydrated NA molecules at the air–water interface, can play an important role in the troposphere.[7] By dividing the present vacuum chamber into photolysis and radical probing sections and connecting them with a differential pumping system, the photolysis of organic molecules on water at atmospheric pressure may be investigated. The photoreactions of NA molecules in the troposphere are considered to be initiated by a spin-forbidden singlet (S$_0$)→triplet (T$_1$) excitation with an absorption peak at around 270 nm.[2] This might cause the cross section of OH radical formation to differ significantly from that in the present 213 nm study. In addition, different photochemical processes such as fluorescence can also occur. Further study is currently in progress.

EXPERIMENTAL METHODS

*LIF measurements*. We developed a new experimental apparatus for the UV laser photolysis of liquid organic molecules and the detection of OH radicals by LIF (Fig. S2). To detect highly reactive OH radicals using LIF, an experimental chamber was evacuated to a background pressure of 1.1 Pa by two dry pumps (Kashiyama, NeoDry15E). The pressure was monitored using a Pirani gauge (Diavac, TRP-11) and a Baratron capacitance manometer (MKS, 722B). Through a fast load-lock door, about 2 ml of neat liquid NA (>98.0%, Tokyo Chemical Industry) was placed on a silicon (111) substrate with dimension of about 35 mm × 45 mm × 700 μm. The substrate was settled on a nickel-plated copper stage controllable with respect to the *x*-, *y*-, and *z*-axes. The temperature of the stage was also controllable using a thermostatic water bath (Thomas, TRL-11LP) and monitored with a type-K thermocouple. A quartz glass Petri dish is not suitable for holding a liquid sample, because it absorbs UV and then emits fluorescence. In comparison, a silicon substrate has the advantage of much smaller noise signals from fluorescence and laser scattering. The typical thickness of liquid NA placed on the substrate was about 810 ± 70 μm, hence photoreactions at the interface between the liquid and substrate can be ignored. The liquid NA was kept at about 295 K to keep its vapor pressure constant during the experiments. Liquid NA has a low vapor pressure of about 0.2 Pa at room temperature, and it can be introduced directly into the medium-vacuum chamber without boiling. In our approach, a quasi-equilibrated and flat liquid surface can be easily achieved compared with experiments using liquid microjets or wetted wheels.[27,28]

As shown in Fig. 1a, a 213 nm Nd:YAG pulsed laser (Lotis TII, LS-2137U-N and VM-TIM, YHG-5) was used for photolysis. This laser was P-polarized and incident on the liquid NA at the angle of 45° to its surface normal. The pulse energy and diameter of the beam were approximately 500 μJ and 4 mm, respectively, which allowed only single-photon absorptio. Desorbed OH radicals were detected by 315 nm LIF via the $A^2\Sigma$ ($v = 1$)–$X^2\Pi$ ($v = 0$) transition. Excited OH ($A^2\Sigma$) was produced by unpolarized pulsed 282 nm radiation from a frequency-doubled dye laser (Lambda Physik, SCANmate 1 with BBO crystal) pumped by a Nd:YAG laser (VM-TIM, VM-45TF5). This probe laser was triggered by a certain delay from the photolysis laser. The pulse energy and diameter of the beam were approximately 50 μJ and 2 mm, respectively. The distance from the liquid surface to the probe laser is represented by $z$, hence the shortest distance between the two lasers ($d$) can be written as $d = z/\sqrt{2}$. A photomultiplier tube (PMT, Hamamatsu, R1104) detected the OH-LIF through a liquid light guide (Lumatec, Series 300) controllable with respect to the $z$-axis, following the set up of McKendrick et al.[29] The distance between the guide and the center of the probe laser was fixed to 5 mm to keep constant the solid angle of OH-LIF detection. Two band-pass filters for wavelengths of 314.5 ± 4.50 nm (Asahi spectra, BPF0313-010) were placed in front of the PMT. The detected OH-LIF was converted to electrical signals by the PMT terminated with a 50 Ω load resistor. Time-resolved fluorescence was recorded by a digital oscilloscope (Iwatsu, DS-5624A). The measurements were typically performed at a repetition rate of 10 Hz using a pulse generator (Quantum Composers, 9428). The minimum detection limit of OH radicals in our set up was estimated from the photolysis of gas-phase AA (>99.7%, Nacalai Tesque) to be approximately less than $3.8(2.6) \times 10^8$ cm$^{-3}$ (see the Supporting Information for details). In the present experiments, AA and NA samples were used as received.

*SFG spectroscopy*. SFG spectra were measured using a laboratory-built HD-VSFG spectrometer, which has been described elsewhere.[20–22] In brief, a femtosecond regenerative amplifier (Coherent, Legend Elite; repetition rate, 1 kHz; center wavelength, 800 nm; pulse width, ~100 fs; average power, 3.0 W) was used to pump two optical parametric amplifiers. The fundamental output from the regenerative amplifier was divided into two. One third of the output was introduced into a white-light-seeded optical parametric amplifier (Coherent, TOPAS-800-fs) to obtain a broadband IR pulse at around 5800 nm (1700 cm$^{-1}$) or 3450 nm (2900 cm$^{-1}$). The remaining two thirds of the output was frequency-doubled to 400 nm by a narrow band second harmonic generator (Coherent, SHBC). The spectrally narrowed second harmonic pulse (FWHM: 8 cm$^{-1}$) was used to pump a white-light-seeded optical parametric amplifier (TOPAS-400-ps-WL) to obtain a visible pulse at 596 nm. The visible and IR pulses were overlapped in a *y*-cut quartz crystal (10 μm thick) to generate a broadband sum-frequency (SF) pulse used as a local oscillator (LO). The transmitted visible, IR, and LO pulses were refocused by a concave mirror onto the sample surface or a *z*-cut quartz surface for reference. The incident angles of the visible and IR pulses were 72° and 60° from the surface normal, respectively. The LO pulse was passed through a 1.5 mm (CH stretching region) or 3 mm thick (C=O stretching region) fused silica plate located between the concave mirror and the sample surface to delay it relative to the visible and IR pulses by ~2.5 ps for the CH stretching region or ~5 ps for the C=O stretching region. The reflected LO and SF pulses from the sample surface were passed through a polarization analyzer to select the polarization of the SFG signal. They were introduced into a polychromator (Horiba Jovin Yvon, TRIAX550; focal length, 550 mm; grating, 1200 grooves/mm) through a prism monochromator (Jasco, CT25-UV), and interfered with each other in the frequency domain. The interfered pulses were detected by a liquid nitrogen-cooled charge coupled device detector (Roper Scientific, LN/CCD-1340/400-EB). The energy of the visible and IR pulses at the sample was ~6 and ~2 μJ/pulse, respectively. The polarization combination for the SF, visible, and IR pulses was SSP (S-polarized SF, S-polarized visible, and P-polarized IR pulses). The height of the sample surface was monitored by a displacement sensor (Keyence, LT8110) and maintained at the same height of the reference *z*-cut quartz surface within an accuracy of 5 μm. The obtained SFG spectra were normalized with those of the reference *z*-cut quartz. For measurements of the C=O stretching region, the sample compartment was purged using dry air to suppress the extinction of the IR pulse by water vapor absorption. Neat NA liquid and aqueous NA solution samples were placed in a Teflon trough (65 mm × 65 mm).

ASSOCIATED CONTENT

**Supporting Information**

The following files are available free of charge.
UV absorption spectrum of neat liquid NA (Fig. S1); photographs of the experimental setup (Fig. S2); typical time-resolved spectrum measured during the 213 nm photolysis of gas-phase AA (Fig. S3); OH A-X (1,0) LIF excitation spectrum measured during the 213 nm photolysis of gas-phase AA (Fig. S4); SFG spectrum in the C=O stretching region of the interface between air and 1.0 mM aqueous NA (Fig. S5); summary of values and errors of the present experiments and literature (Tab. S1); summary of parameters for fitting in Fig. 5B (Tab. S2); summary of parameters obtained by fitting Fig. S5 (Tab. S3); estimation of detection sensitivity of OH radicals.

AUTHOR INFORMATION

Corresponding Authors

*Naoki Numadate

Email: numadate-naoki@g.ecc.u-tokyo.ac.jp

**Tetsuya Hama

Email: hamatetsuya@g.ecc.u-tokyo.ac.jp

ORCIDs

Naoki Numadate: 0000-0001-8992-0812

Yuki Nojima: 0000-0001-9065-6385

Taka-aki Ishibashi: 0000-0003-0000-547X

Shinichi Enami: 0000-0002-2790-7361

Tetsuya Hama: 0000-0002-4991-4044

Notes

The authors declare no competing financial interests.

ACKNOWLEDGMENT

This work was supported by JSPS KAKENHI Grant Numbers JP19K22901, JP20K23361, JP22K05016, and JP22K14644.